\documentclass{aa}
\usepackage{graphicx}
\usepackage{subcaption}
\usepackage{txfonts}

\newcommand{\fink}{{\sc Fink}}

\begin{document}

   \title{Enabling the discovery of fast  transients}

   \subtitle{A kilonova science module for the Fink broker}

   \author{B. Biswas\inst{1}\thanks{\email{biswas@apc.in2p3.fr}},
          E. E. O. Ishida\inst{2}\thanks{\email{emille.ishida@clermont.in2p3.fr}},
          J. Peloton\inst{3}, A. M\"oller\inst{4,5}  
          \and M. V. Pruzhinskaya\inst{2,6} \and R. S. de Souza\inst{7}
          \and D. Muthukrishna\inst{8}
          }

   \institute{Université Paris Cité, CNRS, AstroParticule et Cosmologie, F-75013, Paris, France
         \and
    LPC, Université Clermont Auvergne, CNRS/IN2P3, F-63000, Clermont-Ferrand, France
         \and
    Université Paris-Saclay, CNRS/IN2P3, IJCLab, 91405 Orsay, France
    \and
    Centre for Astrophysics and Supercomputing, Swinburne University of Technology, Mail Number H29, PO Box 218, 31122 Hawthorn, VIC, Australia
    \and ARC Centre of Excellence for Gravitational Wave Discovery (OzGrav), Hawthorn, Victoria, Australia
    \and Lomonosov Moscow State University, Sternberg Astronomical Institute, Universitetsky pr.~13, Moscow, 119234, Russia
    \and Centre for Astrophysics Research, University of Hertfordshire, College Lane, Hatfield, AL10~9AB, UK \and
    Kavli Institute for Astrophysics and Space Research, Massachusetts Institute of Technology, Cambridge, MA 02139, USA
    }

   \date{Received September 15, 1996; accepted March 16, 1997}

  \abstract
   {Large-scale astronomical surveys such as the Zwicky Transient Facility (ZTF) opened a new window of opportunity in the search for rare astrophysical phenomena. Community brokers, such as \fink, have the task of identifying interesting candidates and redistributing them to the community. For the specific case of fast transients, this identification should be done early, based on a limited number of observed photometric epochs, thus allowing it to trigger further observations.}
   {We describe the fast transient classification algorithm in the centre of the kilonova (KN) science module  currently implemented in the \fink\ broker, and we report classification results based on simulated catalogues and real data from the ZTF alert stream.}
   {We used noiseless, homogeneously sampled simulations to construct a basis of principal components (PCs). All light curves from more realistic ZTF simulations were written as a linear combination of this basis. The corresponding coefficients were used as features in training a random forest classifier. 
   The same method was applied to two different datasets, illustrating possible representations of ZTF light curves. The latter aimed to simulate the data situation found within the ZTF alert stream.}
   {Classification based 
   on simulations mimicking ZTF alerts resulted in $69.30 \%$ precision and $69.74 \%$ recall when applied to a simulated test sample, thus confirming the robustness of precision results when limited to 30 days of observations.
   Dwarf flares and point Type Ia supernovae were the most frequent contaminants. The final trained model was integrated into the \fink\ broker and has been distributing fast transients, tagged as \texttt{KN\_candidates}, to the astronomical community, especially through the GRANDMA collaboration.}
   {We show that features specifically designed to grasp different light-curve behaviours provide enough information to separate fast (KN-like) from slow (non-KN-like) evolving  events. This module represents one crucial link in an intricate  chain of infrastructure elements for multi-messenger astronomy, which is currently being put in place by the \fink\ broker team in preparation for the arrival of data from the Vera Rubin Observatory Legacy Survey of Space and Time.}


   \maketitle
%

\section{Introduction}
\label{sec:intro}

The first detection of a gravitational wave (GW) event resulting from a binary neutron star merger  \citep[GW170817,][]{PhysRevLett.119.161101} brought to life a new paradigm in observational astronomy. The detection reported by the LIGO\footnote{\url{https://www.ligo.caltech.edu/}} and Virgo\footnote{\url{https://www.virgo-gw.eu/}} collaborations was accompanied by confirmed signals in gamma, X-ray, ultra-violet, optical, near-infrared, and radio wavelengths \citep{Abbott2017}, and inaugurated a new era of multi-messenger astronomy. From the perspective of optical surveys, the first detected kilonova (KN) came to join gamma-ray burst (GRB) afterglows \citep{2019MNRAS.489.2104T} in the selected group of rare, fast transients possessing a multi-messenger component.  

The idea that the radioactive ejecta from a binary neutron star (BNS) or neutron star-black hole (NSBH) merger provides a source for powering thermal transient emission appeared in the late 1990s~\citep{1998ApJ...507L..59L}. 
The first piece of observational evidence of such a counterpart, lasting days to weeks, was found during the follow-up observations of short GRB130603B~\citep{2013Natur.500..547T}. Later, a KN signature was also suspected in other short GRBs  ~\citep{2018ApJ...860...62G,2019MNRAS.489.2104T}. However, the real interest in KNe arose with GW170817 and its electromagnetic counterparts: short GRB170817A and KN  AT2017gfo~\citep{PhysRevLett.119.161101, Abbott2017}. Detailed modelling of the photometric data of AT2017gfo showed that KNe  are powered by the decay of a wide range of r-process nuclei, and therefore 
this showed that BNS coalescence can be a dominant source of heavy element production in the Universe~\citep{2017ApJ...851L..21V,2017Natur.551...80K}. 
These observations also played a crucial role in locating GW170817, and their association with the same astrophysical source has provided the first piece of direct evidence that at least some short GRBs are associated with BNS mergers. Further observations by Advanced LIGO/Virgo led to the discovery of a few more events, all without any KN signatures~\citep{2021arXiv211103634T}. Whether NSBH and all BNS merges should be accompanied by a KN is one of the exciting puzzles that still needs to be solved.

In parallel, the combined detection of GW and KN signals also provides a new approach for Hubble constant measurements~\citep{1986Natur.323..310S}. The analysis of GW170817 and the redshift of its host galaxy led to a measurement of $H_0 = 74^{+16}_{-8}$~km~s$^{-1}$~Mpc$^{-1}$ (\citealt{2017Natur.551...85A}, see also~\citealt{PhysRevResearch.2.022006}). Such measurements are important since they do not rely on a cosmic distance ladder and they could help in resolving the Hubble constant tension between Type Ia supernovae (SNe Ia) and cosmic microwave background estimates~\citep{2021ApJ...919...16F,2021arXiv211204510R}.

In light of their potential application to such a large range of astrophysical topics, considerable resources have been employed in trying to detect and characterise further KN events. Traditionally, this search is mainly carried out (and planned) in the error boxes of short GRBs and GW alerts~\citep{2018NatCo...9.4089T,2018arXiv181204051M,2020ApJ...905..145K}. However, this approach is limited by the operation schedule of GW detectors, the rapid coordination needed to deploy optical follow-up, telescope availability, and weather constraints, among others. In this context, considerable efforts are already  devoted to optimising the exploitation of wide-field optical survey data for the discovery of KNe independently from a GW detection \citep{Chase2021}. These include analysing data from large facilities such as the Dark Energy Camera \citep{Garcia2020} and the Zwicky Transient Facility \citep[ZTF,][]{Andreoni:2021}, as well as preparing for the arrival of data from the Rubin Observatory in the upcoming years \citep{Lochner2021, hambleton2022}. 

Although photometric detection alone is not able to provide enough information to confirm the classification of a given KN candidate, their estimated rate is expected to be low enough to allow further scrutiny of a significant fraction of high probability transients \citep{Setzer2019}. Thus, rapid and efficient photometric classification is a crucial element to be developed when trying to maximise the number of future confirmed KNe. Nevertheless, the effectiveness of any classification strategy will always be highly correlated with the survey cadence and duration of the target transient. Sources that remain visible for only a few days, such as KNe, pose a greater challenge due to their small number of observed photometric points and the necessity to quickly trigger spectroscopic follow-up. 

Despite these challenges, the  potential for scientific discovery is sufficient enough to motivate investigations focussed on current photometric data. The ZTF survey, covering a large volume of the sky, is an excellent dataset for such searches. Using proprietary data, \citet{Andreoni:2021} developed the ZTF REaltime Search and Triggering  (ZTReST), a complete pipeline devoted to identifying KN candidates and sending the most interesting ones for target-of-opportunity observations. Although no KNe were confirmed during the reported 13 months of operations, the system -- based on filtering, template modelling, and human-in-the-loop feedback -- was able to identify eight fast evolving transients, thus providing evidence of the feasibility of photometric-based approaches. 
To optimise a search of GW optical counterparts, \citet{2020MNRAS.497.1320S} used \textsc{astrorapid} \citep{Muthukrishna2019} --- a classifier tool based on machine learning. Putting all transients in four categories ('KN', 'SN', 'Others', and 'Indistinguishable'), they showed that after a few days of observations, it is possible to rule out the candidates as SNe and other known transients. Recently, \citet{chatterjee2022} used sparse early-time photometry and contextual information to train a temporal convolution neural network with the goal of identifying KN candidates.

In this work, we explore an alternative strategy. Using noiseless simulated light curves, we developed a machine-learning-based classifier capable of identifying fast transients within the ZTF public alert stream, among which a few KN events are to be expected. The algorithm described here is the core of the \fink\ broker \citep{fink} KN classifier-based module and has been reporting candidates since March 2021. These have been publicly forwarded to the whole community, and especially to the GRANDMA telescope network \citep{Antier:2020a}, which coordinates follow-up observations \citep{Antier:2020b, 10.1093/mnras/stac2054}. 

\fink\footnote{\url{https://fink-broker.org/}}\ \citep{fink} is an open-source broker software based on big data and distributive computing techniques. It was specifically designed to face the computational challenges posed by the upcoming Vera Rubin Observatory Legacy Survey of Space and Time\footnote{\url{https://www.lsst.org/}} (LSST) and, as part of its preparatory stages, has been processing ZTF alerts since November 2019.  Along with {\sc ALeRCE} \citep{Forster:2020}, {\sc Ampel} \citep{ampel}, {\sc Babamul}, {\sc Antares} \citep{antares}, {\sc Lasair} \citep{Smith_2019}, and  {\sc Pitt-Google}, it is part of the broker ecosystem which will distribute LSST alerts to the community throughout the next decade. 

This work is organized as follows. 
In section \ref{sec:data} we describe the dataset used to generate the basis vectors and the datasets used for training and testing our classifier.
In Section \ref{sec:method} we dive deeper into the method for fitting and the classification of light curves, 
while Section \ref{sec:results} describes the results of our approach for the classification of simulated KN events. Section \ref{sec:fink} gives details on the candidates found within the real ZTF alert stream. We present our conclusions in Section \ref{sec:conclusions}.
All codes used to produce the results presented here are publicly available\footnote{\url{https://github.com/b-biswas/kndetect}}.

\section{Data}
\label{sec:data}

We used two distinct datasets: a set of ideal simulations (hereafter \texttt{perfect\_sims}) used to construct a basis for feature extraction, and a set of ZTF-like simulations (hereafter \texttt{ztf\_sims}) used to demonstrate the performance of our method. The \texttt{perfect\_sims} consisted of light curves observed through the LSST $[g,r]$ broad-band filters with typical noise photon count of $10^4$, no galactic extinction, and without intrinsic magnitude smearing (\texttt{PERFECT} light curves according to the SNANA simulator\footnote{\url{https://github.com/RickKessler/SNANA/blob/master/doc/snana_manual.pdf}}, \citealp{Kessler2009}), and considering a uniform cadence of 2 days. 
This set is composed of 1000 events generated from KN  template models \citep{kasen2017, Kessler2019} and 1000 events generated from non-KN templates. Among the latter, we included 125 simulations for each class: SN Ia-91bg, SN II-NMF, SN IIn, SN Ia, SN Iax, and  SN Ibc-MOSFIT from \citealp{Kessler2019}, as well as SN Ib and SN II  from \citealp{Vincenzi2019}.  
The non-KN subset contains the light-curve behaviours expected from intermediate-term transients (SN-like) while the KN subset provides the contrasting shape of short-term transients. 
Although this module was constructed aiming at ZTF real data, we used LSST filters due to their easy availability within SNANA and the fact that they show similar enough transmissions to the ZTF filters, thus being able to retain the overall shape of the light curves. 
Since at this stage we are not interested in details of the light-curve shape, this was a good enough approximation (all light curves generated with SNANA were given in terms of its arbitrary flux unities, \texttt{FLUXCAL}). 
In Section \ref{sec:method} we show how this diversity was used to create a meaningful parameter space that enables one to separate short ($\sim$ 1 week) and intermediate (a few month-long) transients in the ZTF alert stream. 

The \texttt{ztf\_sims} dataset was constructed by merging two different simulations. The first set was presented in \citet{Muthukrishna2019}. It contains 38 000 non-KN and 4568 KN light curves. 
They were also generated using SNANA, as well as models and rates from the Photometric LSST Astronomical Classification Challenge \cite[PLAsTiCC,][]{Kessler2019} and observational characteristics, such as cadence and observation conditions, from the Mid Scale Innovations Program (MSIP) survey at the ZTF \citep{Bellm2014}, using the two filters available in the ZTF public survey [$zg$, $zr$].  
For further details on the simulation properties, readers can refer to \citet{Muthukrishna2019}. The redshifts in these models range up to $\approx 0.1$ for KN and $\approx 1.2$ for non-KN models.
The second set (hereafter \texttt{GRANDMA}) holds a total of 1000 simulated KN light curves described in \cite{Stachie:2019nae}, and it represents a more recent cadence strategy of ZTF. Thus, the combined \texttt{ztf\_sims} dataset contains light curves representing two different cadences used in the ZTF survey. This dataset represents the state of the art in what concerns publicly available information for simulating the complete sky as observed by ZTF. Results reported on this simulated dataset (Section \ref{sec:results}) will certainly get closer to the ones obtained from live real data (Section \ref{sec:fink}) if new template models for simulations are made available.

In summary, the \texttt{ztf\_sims} dataset holds 5568 KN and 38 000 non-KN objects observed in two ZTF filters. We used this to quantify the efficiency of our method when applied to the ZTF alert stream.
In Section \ref{sec:method} we describe how this dataset was further divided to train and test the classifier.

\section{Methodology}
\label{sec:method}

Our approach is divided into three steps: template generation (Section \ref{subsec: Template Generation}), feature extraction, and classification (Section \ref{subsec: Feature generation and classification}). We used the \texttt{perfect\_sims} set in order to construct three basis functions over which all the \texttt{ztf\_sims} will be projected. The projection coefficients were then used as features and submitted to a random forest classifier. We give details of each step in the subsequent sub-sections. 

\subsection{Template generation}
\label{subsec: Template Generation}

\begin{figure}[!t]
\centering
  \includegraphics[width=\linewidth]{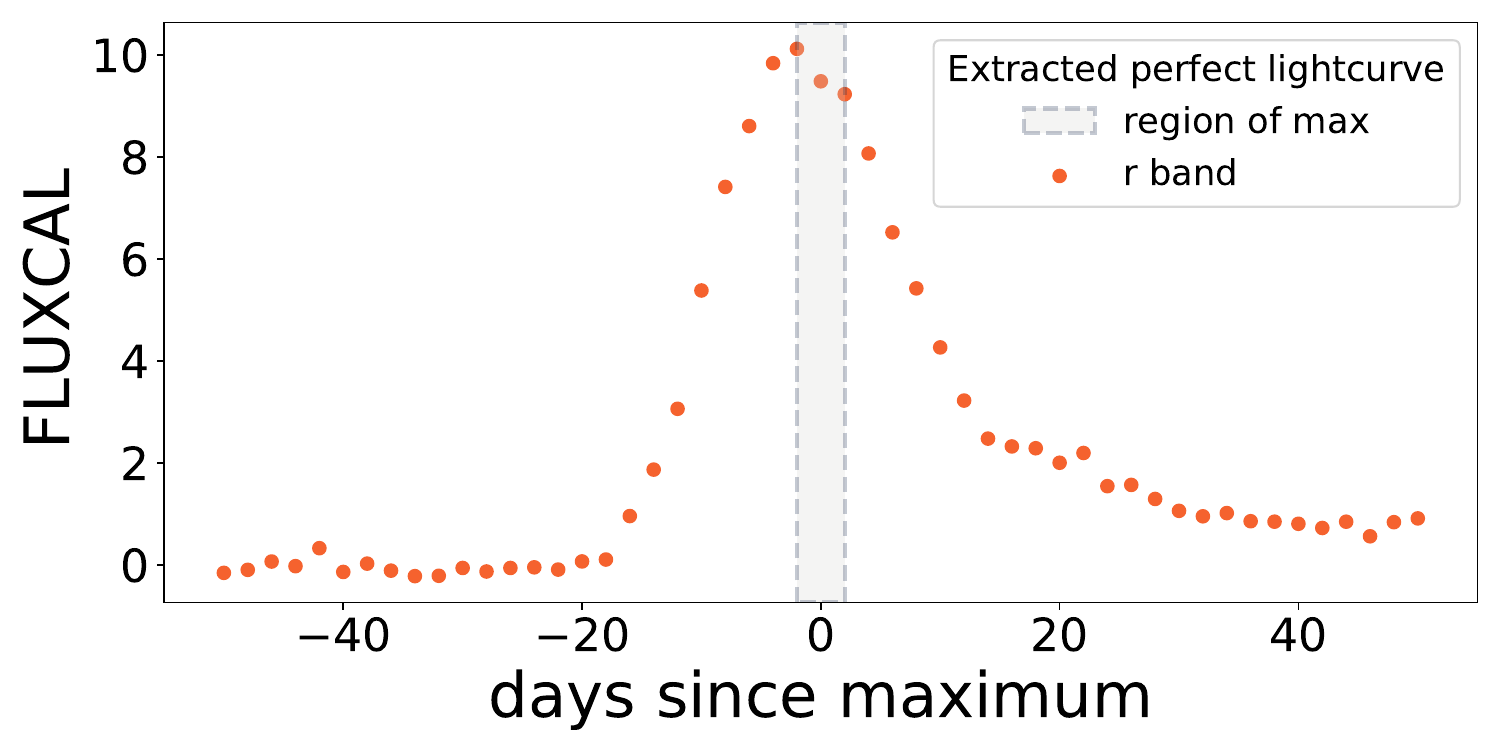}
  \caption{Example of the light curve from \texttt{perfect\_sims} in the $r$ $band$ with maximum brightness anchored at index 24. The grey region marks the possible anchor positions, equivalent to indexes 24, 25, and 26 in the vector representation described in Section \ref{subsec: Template Generation}.}
  \label{fig:lc_ex}
\end{figure}

We represented each light curve in \texttt{perfect\_sims} as a vector of 51 elements, $\bar{v}$, with each element holding the flux measurement in a given epoch, from -50 to +50 days in bins of 2 days. 
Maximum flux was randomly positioned among indexes [24, 25, 26], consequently, the maximum measured flux is at $0 \pm 2$ days. 
We allowed for this slight shift during anchoring so as to take into account the fact that the maximum observed flux may not be the true maximum due to the telescope  cadence.
In cases where the light curve does not hold enough epochs, we filled the empty elements of $\bar{v}$ with zeros. 
Figure \ref{fig:lc_ex} shows an example of an SN Ia-91bg at redshift $z\approx0.5$ with the day of maximum anchored at index 24. All light curves in the \texttt{perfect\_sims} set were subject to this centralisation procedure (each filter was treated independently). 
We then constructed a data matrix stacking all vectors $\bar{v}$ from both bands and classes ( $\mathcal{D}_{\rm template}$ - 4000 rows, 51 columns). 

\begin{figure}[!tb]
  \centering
  \includegraphics[width=.9\linewidth]{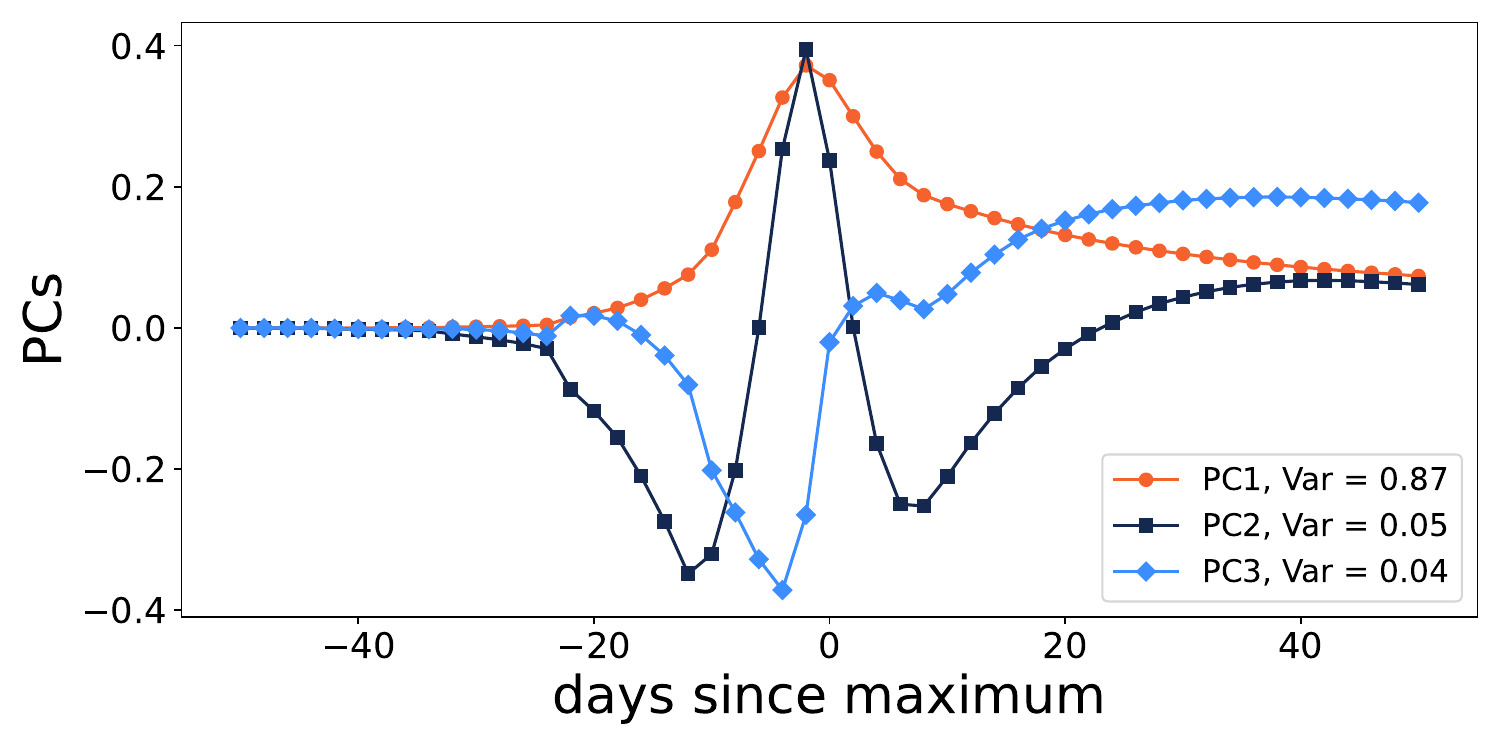}
  \caption{Principal components generated from the \texttt{perfect\_sims} dataset. The plot shows the first three PCs with the highest variance as generated from PCA on ($\mathcal{D}_{\rm template}$).}
  \label{fig: PCs}
\end{figure}

The matrix was then submitted to principal component analysis (PCA) using the scikit-learn library \citep{scikit-learn}. 
Figure \ref{fig: PCs} shows the first three components with the largest fraction of total variance, obtained from  $\mathcal{D}_{\rm template}$. We searched for the minimum number of PCs that encodes enough information to allow for light-curve reconstruction while, at the same time, restricting the number of degrees of freedom necessary for feature extraction. We chose to keep the first three PCs, as together they hold 96\% of the total data variance.

\begin{figure}[!t]
 \centering
  \includegraphics[width=.9\linewidth]{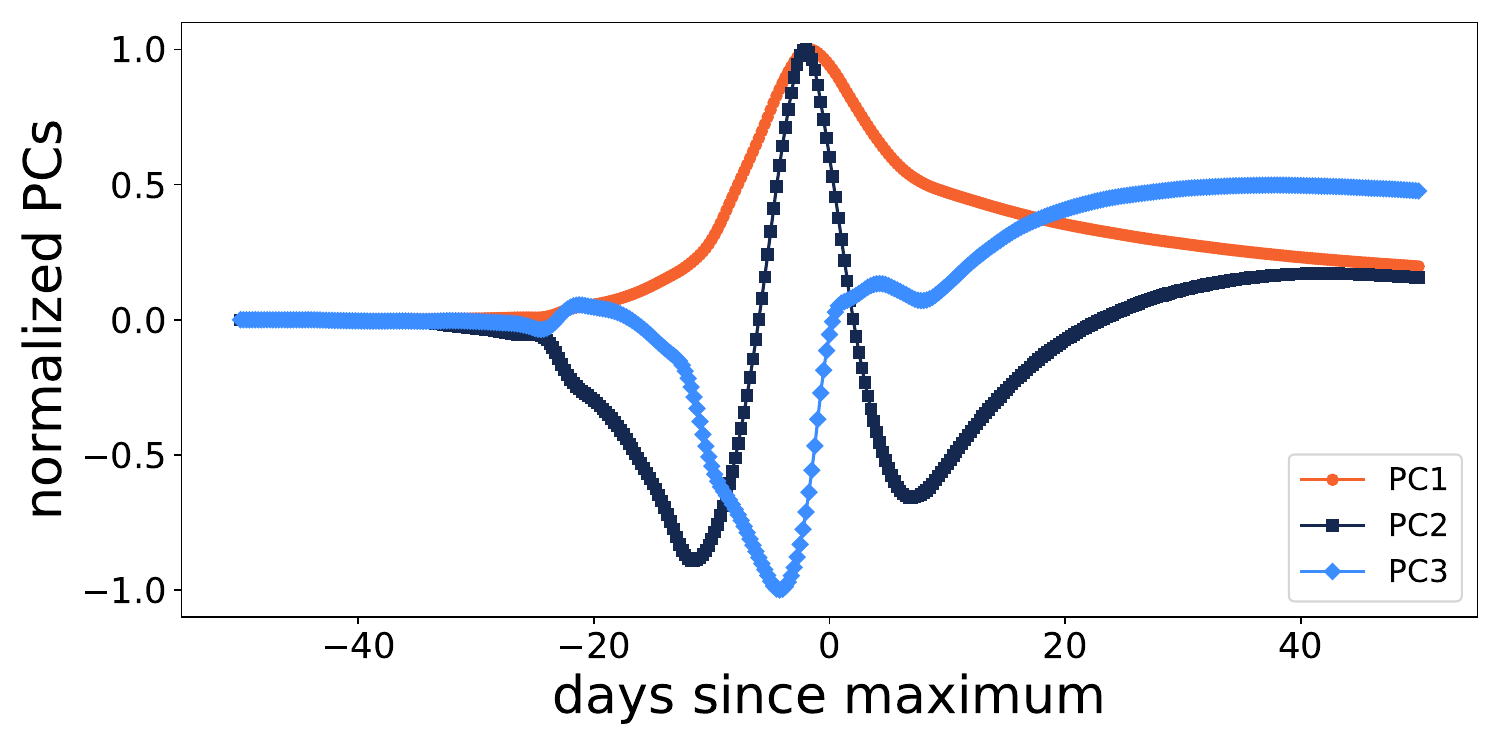}
  \caption{Final interpolated versions of PCs in Fig \ref{fig: PCs}. 
  }
  \label{fig:PCs fit}
\end{figure}

We aim to use the projections of each light curve in these three chosen components as features for further machine-learning analysis. 
While we used the same set of basis vectors for both bands of ZTF data, we expect the projections to take into account the variability between the bands.
Thus, it was necessary to allow for more frequent cadences to be coded within each light-curve vector, $\bar{v}$, without loss of information. We achieved this by performing a quadratic spline fit in each component and then representing it in bins of 0.25 days (6 hours), which results in components of 401  dimensions.
Figure \ref{fig:PCs fit} shows the three components, normalised to amplitude one, as they appear in this final base $\mathcal{P} = [p_1, p_2, p_3]$. In what follows, the same set of PCs was used to extract features from both ZTF passbands.

\subsection{Feature extraction}
\label{subsec: Feature generation and classification}

For each filter in the measured light curve ($\bar{l}$), the flux values were normalised by the maximum observed flux($f_{\rm max}$), thus allowing the reconstruction to focus on the light-curve shape instead of absolute brightness. 
This information was later included by adding $f_{max}$ as a feature to the classifier. 
Further, to minimise noise effects, we only considered objects with $f_{\rm max} > 200$.
This $f_{max}$ cutoff corresponds to a magnitude of $\approx 21.75$, which is close to the limiting magnitude of ZTF \citep[][, Figure 6]{Bellm_2019}.

Each light curve from \texttt{ztf\_sims}, in a given filter, was reconstructed as a vector (hereafter {predicted} light curve, $\bar{l}_p$) of 401 elements, with each element corresponding to an epoch from -50 to + 50 days in steps of 6 hours. 
We aligned $f_{\rm max}$ at the centre of this vector:

\begin{equation}
    \bar{l}_p = \sum^{k}_{i = 1}c_{i}*p_{i},
    \label{eqn: prediction}
\end{equation}
where $c_{i}$  is the coefficient of $i^{th}$ base element, $p_{i}$,  and $k = 3$ is the number of elements in $\mathcal{P}$. The best-fit values for $c_i$  parameters were found by minimising the loss function:
\begin{eqnarray}
    loss & = &\sum_{i}^N\frac{(l_{p,i} - l_{i})^{2}}{\sigma _{i}^{2}} + \left[\sum_{k=1}^3 c^2_k - c^2_1 H(c_1)\right]\frac{f_{\rm max}^2}{\sigma_{f_{\rm max}}^2},
    \label{eqn: loss penalty}
\end{eqnarray}

\begin{figure}[!t]

    \begin{subfigure}[b]{0.55\textwidth}
       \includegraphics[width=.9\linewidth]{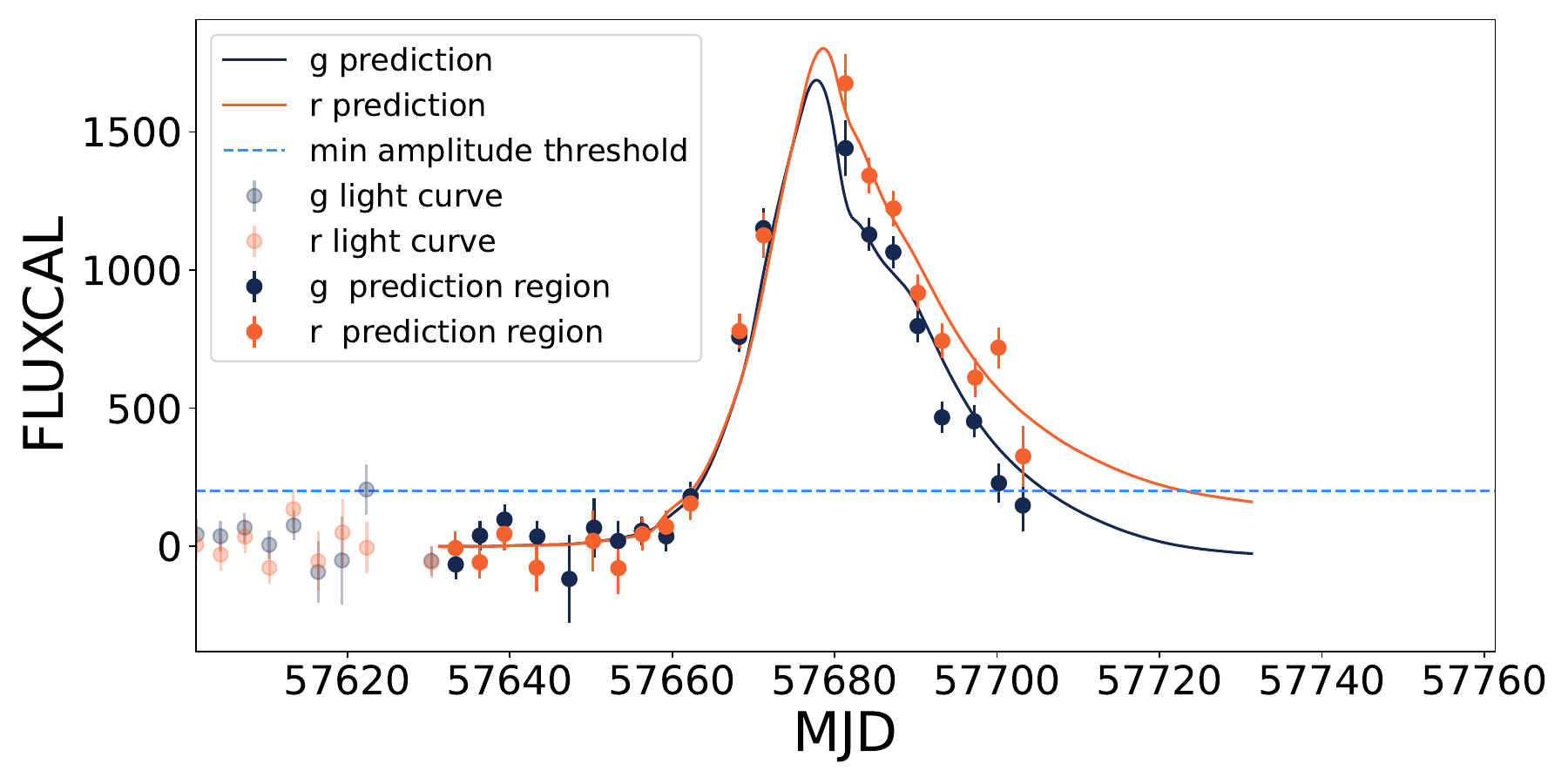}
       \caption{SN Ia event at $z \approx 0.13$}
       \label{fig: fit 1} 
    \end{subfigure}

    \begin{subfigure}[b]{0.55\textwidth}
       \includegraphics[width=.9\linewidth]{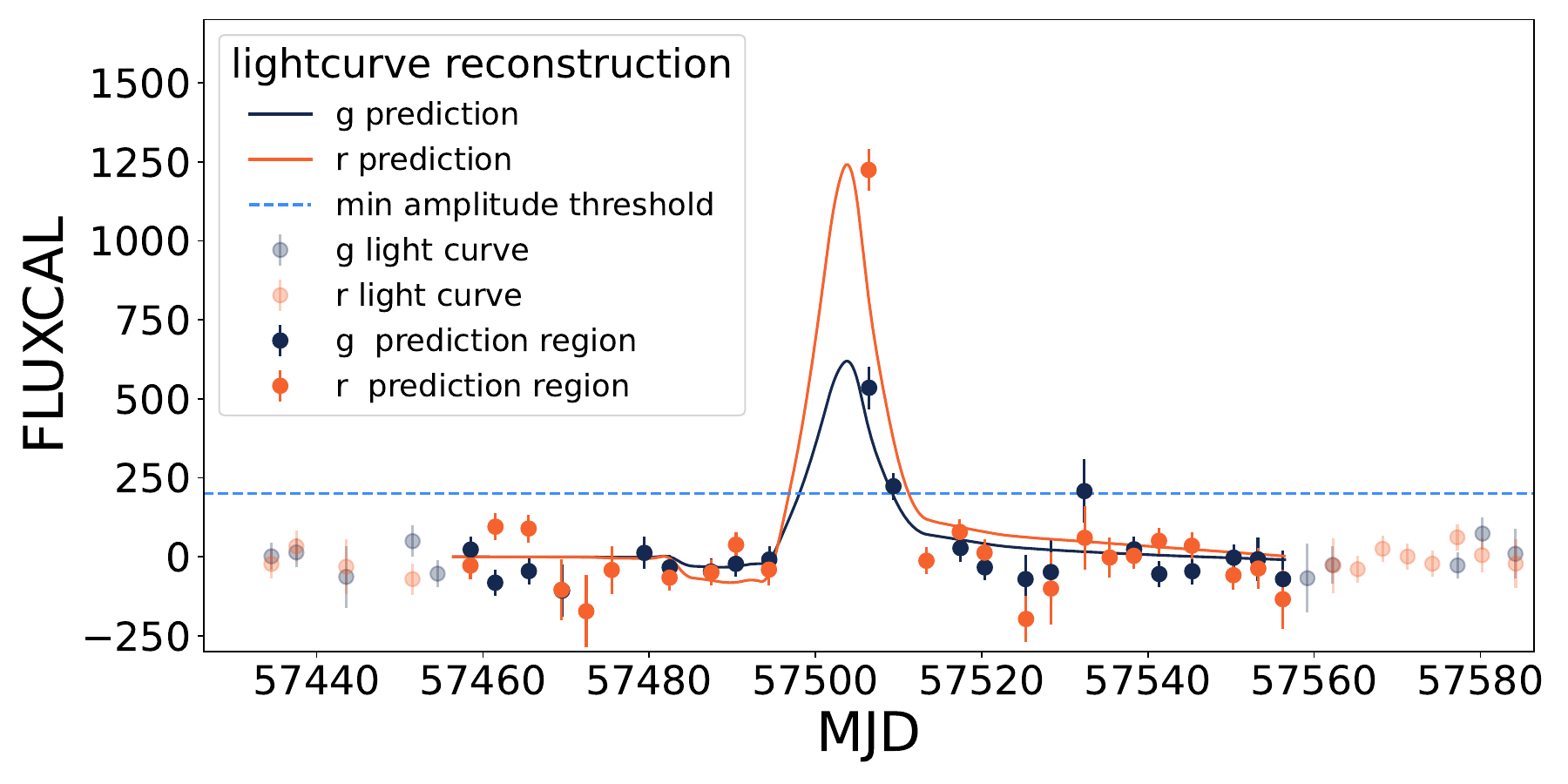}
       \caption{KN event at $z \approx 0.02 $}
       \label{fig: fit 2} 
    \end{subfigure}

  \caption{Light curves from the \texttt{ztf\_sims} dataset. Points correspond to measurements from SN Ia (top) and kilonova (KN, bottom) events, observed in the two ZTF filters. Dark points mark the measurements used for feature extraction, while transparent ones lie outside the prediction region ($\pm 50$ days since maximum). Full lines show the reconstructed light curve and the dotted horizontal light blue line marks the minimum flux threshold. 
  }
  \label{fig: SN fit}
\end{figure}

\noindent where $l_{p,i}, l_i$, and $\sigma_i$ are the predicted flux, the measured flux, and the flux  error at the $i$-th element of $\bar{l}$, respectively, $\sigma_{f_{\rm max}}$ is the flux error associated with $f_{\rm max}$, $H(x)$ is the Heaviside function, and $N$ is the total number of photometric observations. The first term in Equation \ref{eqn: loss penalty} accesses the goodness of fit of the final reconstruction. 
The second term (regularisation term) was introduced to guarantee a stable reconstruction. It ensures that $c_1$ has positive values and only allows relatively smaller magnitudes for $c_2$ and $c_3$, since they represent only a small percentage of the total variance (5\% and 4\%, respectively). 
The scaling factor in the regularisation term ensures that the two terms in equation \ref{eqn: loss penalty} are comparable and it prevents over-fitting in cases where the light curve has a small number of points.
We show two examples of light-curve reconstruction in Figure \ref{fig: SN fit}.

\begin{figure*}[!t]
  \includegraphics[width=\linewidth]{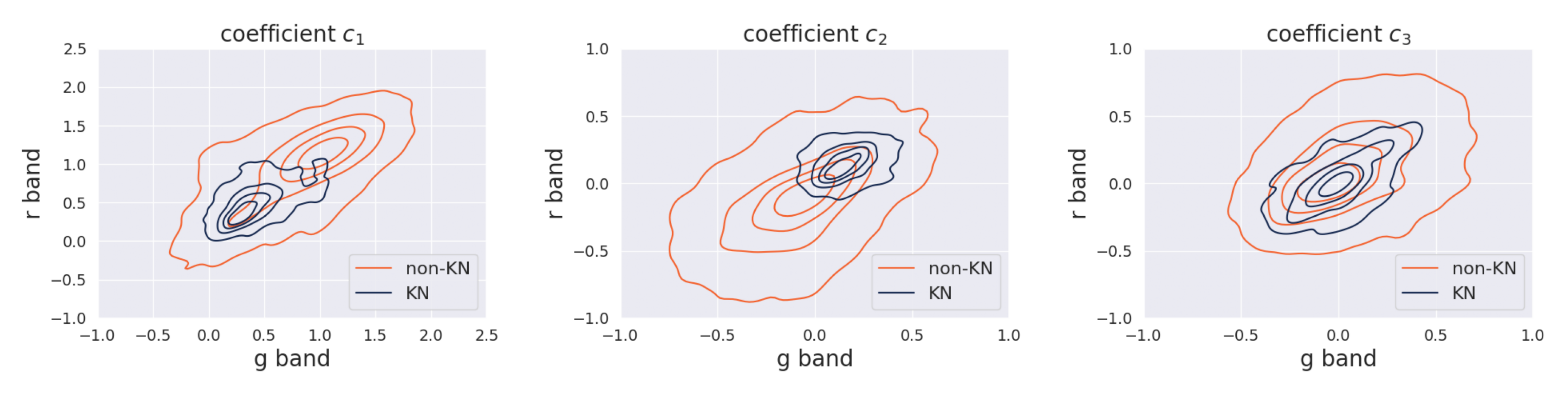}
  \caption{Distribution of $\mathcal{F}_{\rm test}$ (Section \ref{subsec:class}) events in the features' space. Orange contours denote  non-KN and blue contours mark KN events. The panels show distributions in different filters for $c_1, c_2$, and $c_3$ (coefficients associated with the basis vectors $p_1, p_2$, and $p_3$ respectively). }
  \label{fig:band correlation}
\end{figure*}

\begin{figure*}[!t]
  \includegraphics[width=\linewidth]{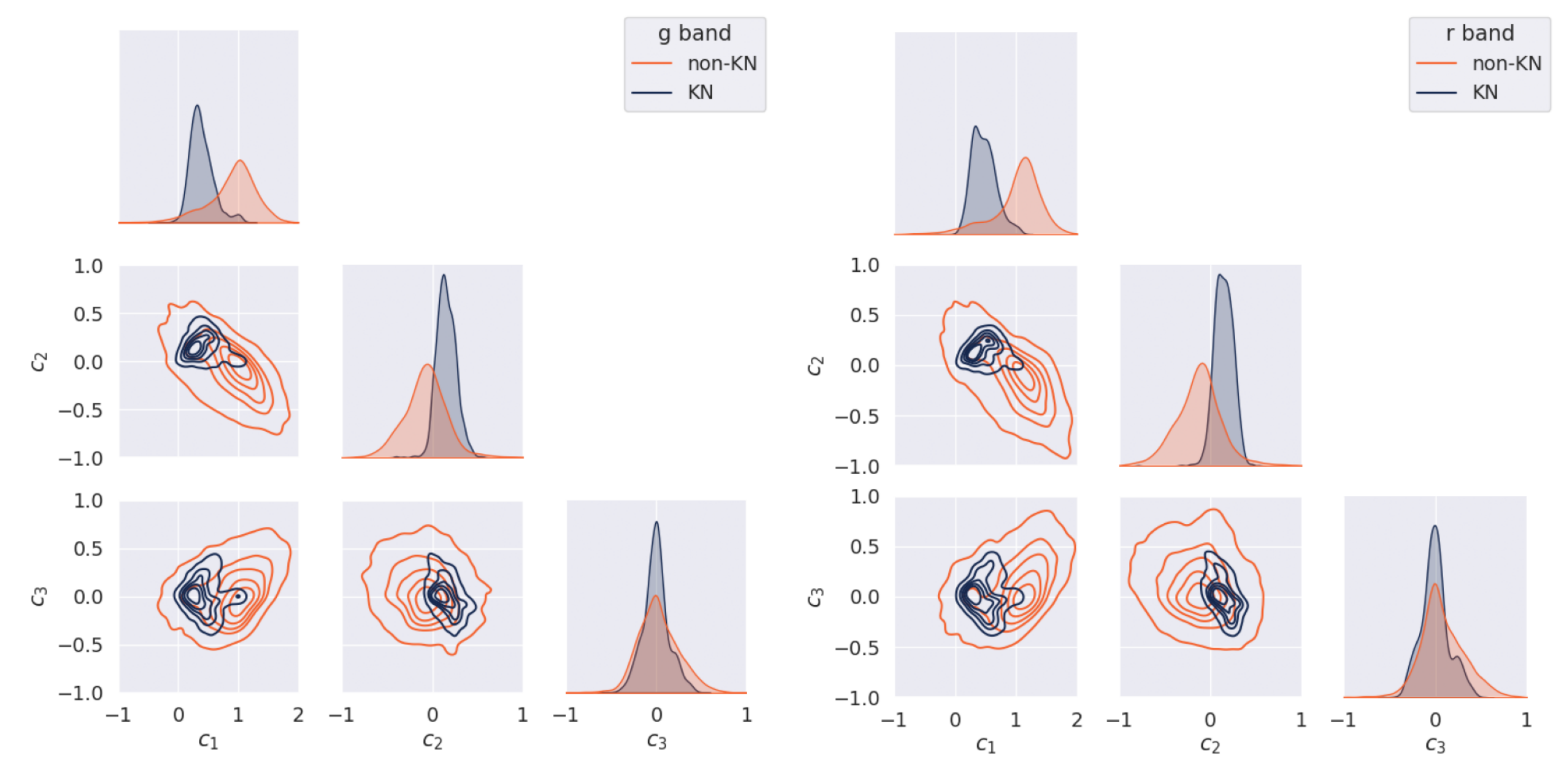}
  \caption{
  Distribution of $\mathcal{F}_{\rm test}$ (Section \ref{subsec:class}) events in the features space. Orange contours denote  non-KN and blue contours mark KN events. The panels show distributions in different filters for $c_1, c_2$, and $c_3$ (coefficient associated with basis vectors $p_1, p_2$, and $p_3,$ respectively).
  }
  \label{fig:coeff correlation}
\end{figure*}

The best-fit parameters for $c_i$  (Equation \ref{eqn: loss penalty}) were computed using the \texttt{SciPy} library \citep{2020SciPy-NMeth} with the default optimisation method \texttt{L-BFGS-B}, described in \citet{53712fe04a3448cfb8598b14afab59b3} and \citet{741068676f0d4748ba518263a9ca1363}. For each object, we concatenated the parameters corresponding to filters  $[g,r]$.

The final feature set per object and per filter was formed by the three best-fit parameter values, $c_i$, $f_{\rm max},$ and a measurement of the residual of the fit,

\begin{eqnarray}
    {\mathcal R} & = &\sqrt{\frac{1}{N}\sum_{i}^N\frac{(l_{p,i} - l_{i})^{2}}{\sigma _{i}^{2}}}.
    \label{eqn: residuo}
\end{eqnarray}

By submitting all objects in \texttt{ztf\_sims} to this feature extraction procedure, we obtained a features matrix, $\mathcal{F}$, of 38221 rows and ten columns (five for each ZTF filter). Figures \ref{fig:band correlation} and \ref{fig:coeff correlation} show the distribution of objects in light-curve-feature space. 
The plots show that KN events occupy a more restricted region of the parameter space when compared to non-KN sources. This property, when combined with the remaining parameters (maximum flux and residual) allows the feature space to  separate KN from non-KN events in this simulated scenario. 
In particular, we notice from Figures \ref{fig:band correlation} and \ref{fig:coeff correlation} that KN events tend to have lower values of $c_1$ and positive values of $c_2$, while non-KN events have a higher contribution from $c_1$ and negative values for $c_2$. 
This behaviour is expected given the nature of generated PCs.

\begin{figure}[!t]
  \centering
  \includegraphics[width=110pt]{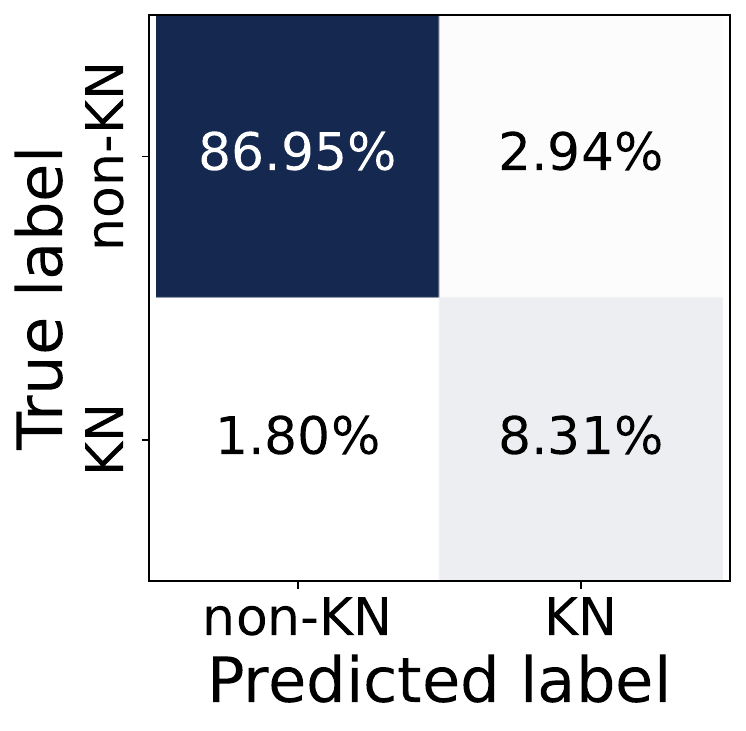}
  \caption{Confusion matrix from the long light-curve baseline scenario. 
  }
  \label{fig: confusion matrix full light curve}
\end{figure}

\subsection{Classification}
\label{subsec:class}

We randomly divided $\mathcal{F}$ into two subsets: train and test.
The training sample $\mathcal{F}_{\rm train}$ contained 2794 KN and 16712 non-KN elements, while the test sample, $\mathcal{F}_{\rm test}$, enclosed 1892 KN and 16823 non-KN events. 

Given the low fraction of KN events and the binary nature of our classification task, a simple random split of the available simulated data into training and test samples would require some fine-tuning of the probability threshold used to define KN candidates. Moreover, this would also result in a large size for the final training model, which is not ideal for operations within the broker. Aimed at circumventing both issues, we followed the active learning strategy \citep{settles2012} based on the uncertainty sampling strategy  described in \citet{leoni2022} to build an informative and balanced training sample. Only $\mathcal{F}_{\rm train}$ was used in the active learning loops. The initial training sample, $\mathcal{F}_{\rm ini\_train}$, contained ten objects (five KNe and five non-KNe) and the system ran through 1490 iterations, thus resulting in a final training sample, $\mathcal{F}_{\rm final\_train} \in \mathcal{F}_{\rm train}$,  of 1500 objects equally balanced between the two classes. Allowing a further increase in the number of objects in training did not improve the classification results. Once this optimal training sample was built, we used it to train a random forest classifier containing 30 decision trees and a maximum depth of 42, using the \texttt{scikit-learn} implementation \citep{scikit-learn}.  Classification results reported below used a probability threshold of 0.5 and were obtained by applying the resulting model to the completely independent $\mathcal{F}_{\rm test}$ sample.

We evaluated the performance of the classifier in terms of   

\begin{eqnarray}
    {\rm Precision} & = & \frac{TP}{TP + FP}
    \qquad {\rm and} \label{eqn: precision}\\
    {\rm Recall} & = & \frac{TP}{TP + FN},
    \label{eqn: recall}
\end{eqnarray}

\noindent where $TP, FP$, and $FN$ stand for true positive, false positive, and false negative, respectively. In the context of this work, the KN class is considered positive, and the non-KN class is considered negative. Thus a $FP$ would be a non-KN classified as KN, and so forth. Precision is also known as purity, while recall is associated with the efficiency of a given classifier.


\section{Results}
\label{sec:results}

\begin{figure}
  \centering
  \includegraphics[width=\linewidth]{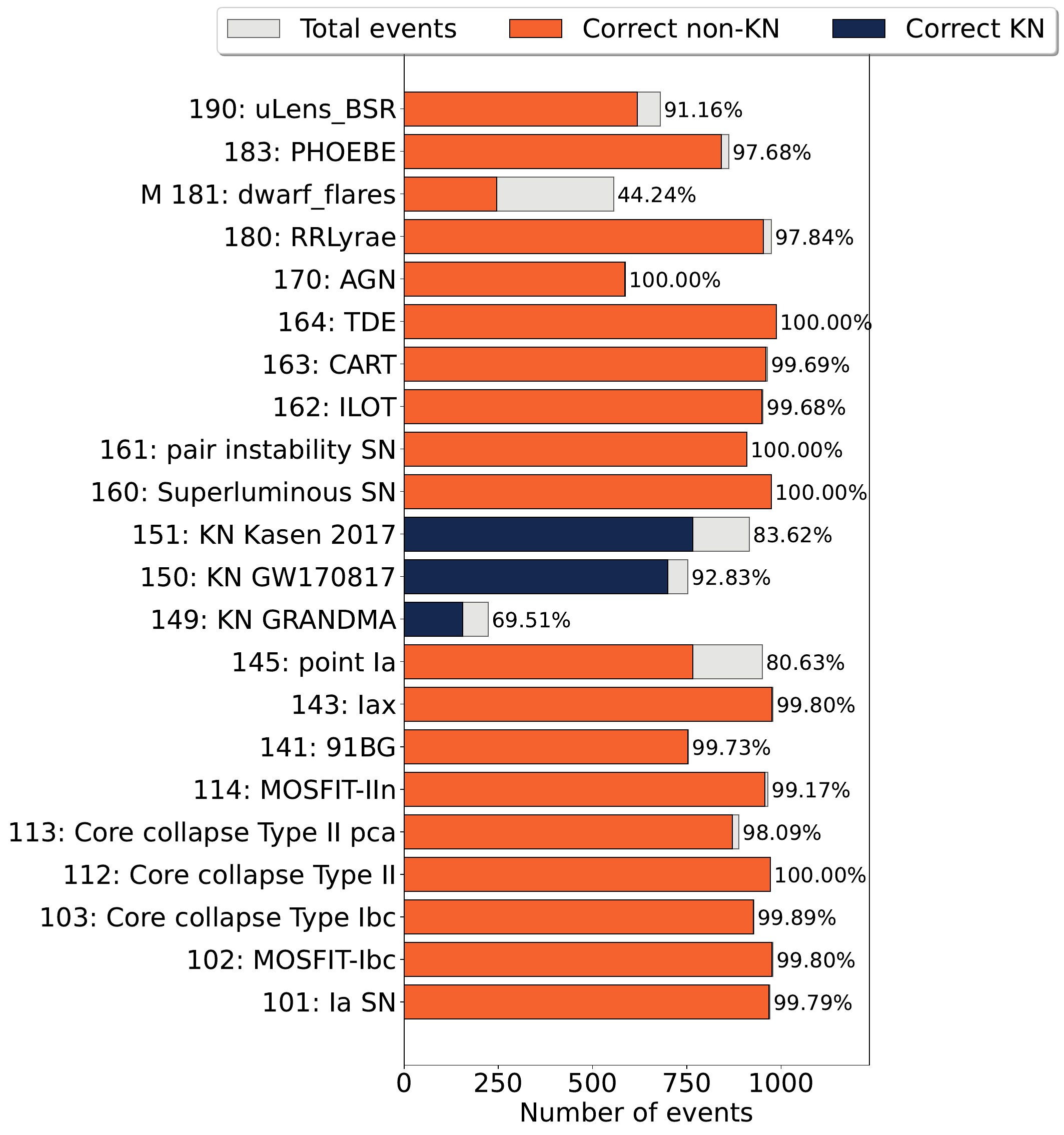}
  \caption{Classification results for the long light-curve baseline scenario. The classifier was trained following a binary configuration of the training set (KN vs non-KN).  We show the results separated by class in order to allow for a better understanding of the contaminants. The plot shows the SNANA code and class names as a function of the number of elements in the test sample. Each bar corresponds to the total number of objects of each class in the test sample. Orange bars denote non-KN and blue bars correspond to KN models. The number to the right of each bar reports the percentage of events correctly classified for each class.
  }
  \label{fig: Contamination start for complete light curves}
\end{figure}

\begin{figure}[!t]
  \includegraphics[width=\linewidth]{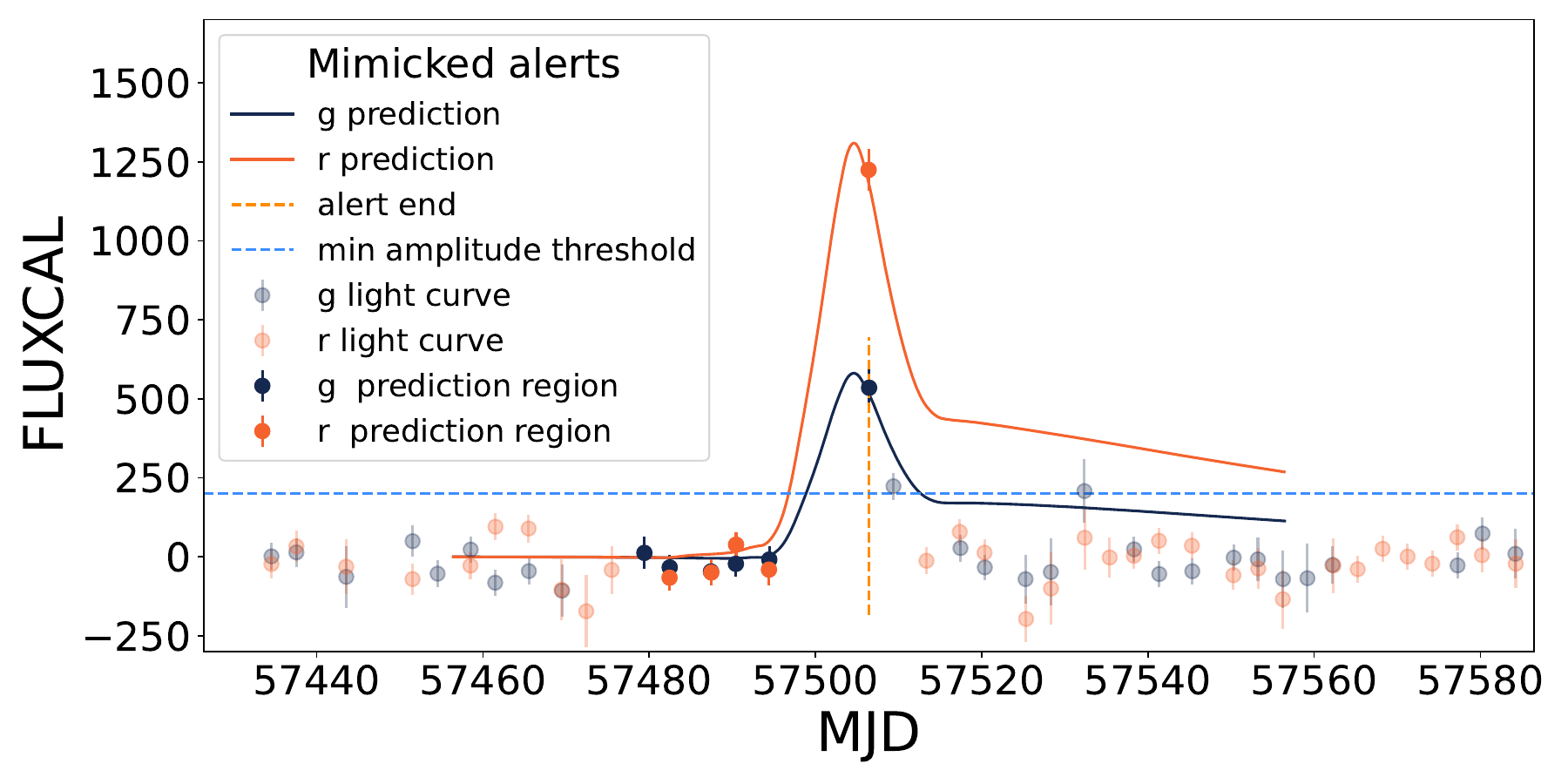}
  \caption{Example of a light-curve fit using only 30 days of measurements (medium baseline) for a KN of class 151 (KN Kasen 2017) at a redshift $z=0.02$. The horizontal-dashed light blue line shows the minimum flux threshold, while the vertical-dashed  orange line represents the last epoch of data considered for the fit. Dark points show photometric measurements used for feature extraction and the fainter points represent the remaining points in the light curve. The full lines denote the reconstructed light curves.  
  }
  \label{fig: KN partial fit}
\end{figure}

\begin{figure}[!t]
  \centering
  \includegraphics[width=110pt]{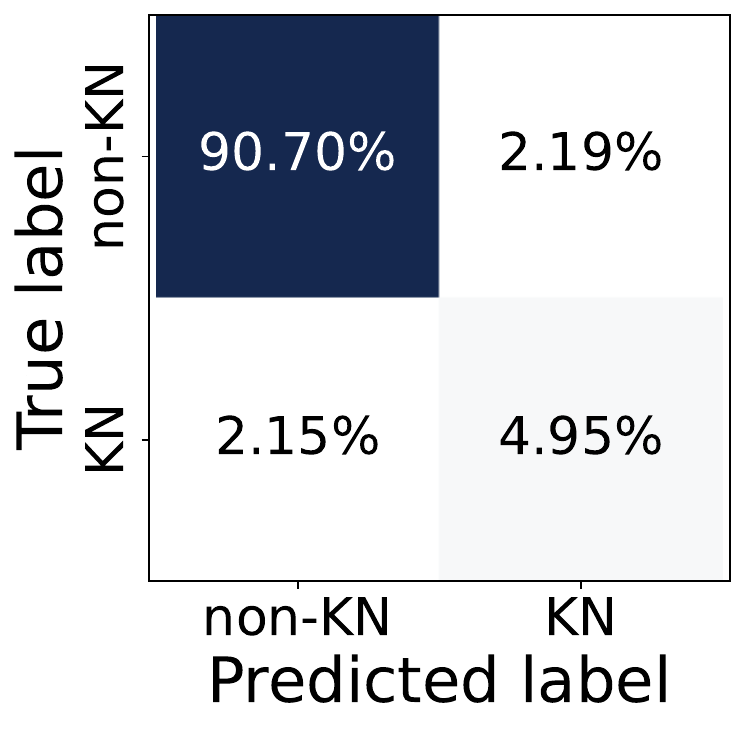}
  \caption{Confusion matrix obtained from the medium light-curve baseline classification experiment.
  }
  \label{fig: confusion matrix partial light curve}
\end{figure}

Using the procedure described in Section \ref{sec:method}, we report results from two distinct scenarios: long (LSST-like duration, Section \ref{subsec: Full light curve}) and medium  (ZTF-like duration, Section \ref{subsec:partial_lc}) light-curve baselines. Both scenarios include non-detection  points (\texttt{PHOTFLAG}$ = 0$ in SNANA). The long light-curve scenario  is an important test to understand how our classifier works when complete information is available. For example, this scenario is less likely to contain objects whose signal is  within the boundaries of the light curve. However, it does not tell us how it is supposed to perform when applied to the ZTF alert stream. To mimic the data situation faced by \fink\  when processing ZTF alerts, we generated  medium light curves lasting for a maximum of 30 days (randomly selecting a 30-day interval for each of the long light-curve baseline objects) and submitted them to the same feature extraction and classifier procedure. We also provide a brief report on the experience of advertising KN candidates produced by this module via the \fink\ broker (Section \ref{sec:fink}).

\subsection{Long light-curve baseline}
\label{subsec: Full light curve}

The random forest classifier trained in $\mathcal{F}_{\rm final\_train}$ was applied to $\mathcal{F}_{\rm test}$, resulting in a precision of $73.87 \%$ and a recall of $82.19 \%$ for the long baseline dataset (Figure \ref{fig: confusion matrix full light curve}). Figure \ref{fig: Contamination start for complete light curves} shows a complete picture of the classification results according to the different classes within $\mathcal{F}_{\rm test}$. We clearly see that a significant part of the contamination comes from dwarf flares, which are also very sharply peaked. Although there is significant misclassification of dwarf flare events, we still notice that the classifier is able to correctly identify $\approx 44\%$ of such events by exploiting the correlations among the features.
The figure also shows that a significantly lower fraction of the  \texttt{KN GRANDMA} objects are available for the classification case. This is a consequence of the lack of forced photometry in these simulations. Since we require a minimum of two points per filter, one of them with \texttt{FLUXCAL} above 200, the lack of baseline points marking a non-detection significantly decreases our ability to extract features from these events.

\begin{figure}[!t]
  \includegraphics[width=\linewidth]{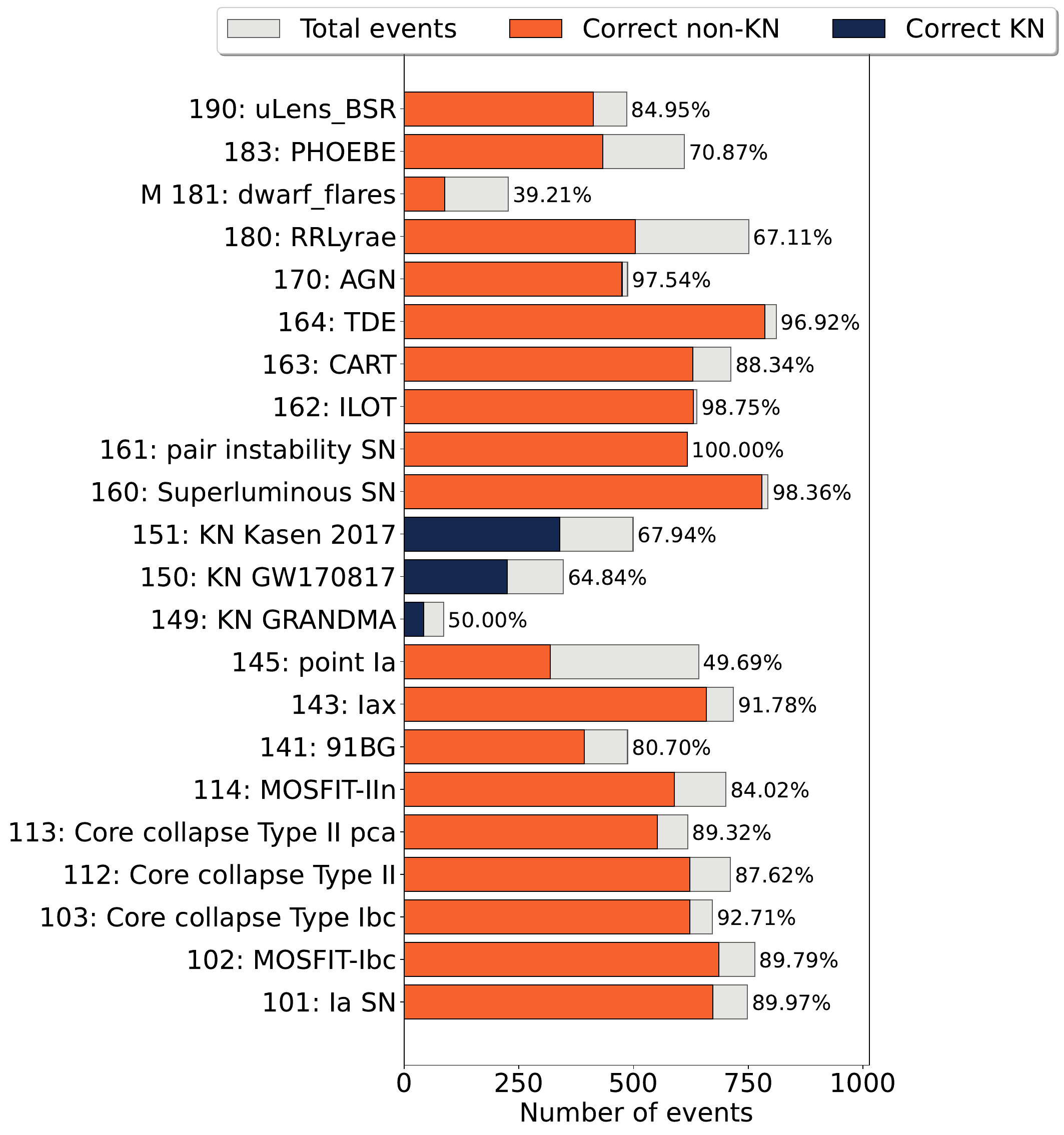}
  \caption{Classification results for the medium light-curve baseline scenario. The description of each element is equivalent to those described in Figure \ref{fig: Contamination start for complete light curves}.}
  \label{fig: Contamination start for partial light curves}
\end{figure}

\subsection{Medium light-curve baseline}
\label{subsec:partial_lc}

In order to mimic the data situation encountered by \fink\ while  processing ZTF alerts,  we generated a more challenging dataset using only some of the light-curve information. For each light curve from \texttt{ztf\_sims} represented as the data vector $\bar{l}$ (Section \ref{sec:method}), we randomly selected a time bin with flux $>= 200$, $t_{\rm now}$, and discarded all elements of $\bar{l}$ corresponding to epochs outside the interval  $[t_{\rm now} - 30, t_{\rm now}]$. The resulting $\bar{l}$ then underwent the same procedure described in Section \ref{subsec:class}. Figure \ref{fig: KN partial fit} shows an example of a KN light curve at redshift $z=0.04$ and its reconstruction using only a 30-day interval. 
A total of 26672 (2380 KN and 24292 non-KN) light curves survived this pre-processing (1448 KN, 12099 non-KN in $\mathcal{F}_{\rm train}$ and 932 KN, and 12193 non-KN in $\mathcal{F}_{\rm test}$). As expected, the fraction of KN to non-KN in this scenario is significantly lower than in the long light-curve case. This is a direct consequence of the shorter signal region from KN events, which results in most of the random choices of $t_{\rm now}$ resulting in medium light curves with $f_{\rm max} < 200$. 

The training sample described above was submitted to the same active learning procedure described in Section \ref{subsec: Full light curve}. The final training sample containing 1500 objects was given as input to a random forest classifier containing 30 decision trees. The resulting confusion matrix as shown in Figure \ref{fig: confusion matrix partial light curve} corresponds to a precision of $69.30 \%$ and a recall of $69.74 \%$, obtained from the medium baseline dataset (30 day light curves). 
As expected, the lower information content present in the medium light-curve baseline resulted in a significant decrease in performance, meaning that a larger number of KNe were missed. Nevertheless, the decrease in precision was much smaller, indicating that even in the medium light-curve scenario the purity of the sample classified as KN remained under control.  Figure \ref{fig: Contamination start for partial light curves} shows a more detailed version of the classification as a function of classes. The main contaminants (non-KN events classified as KN) continue to be fast-rising transients, such as dwarf flares and point Ia SNe, indicating that the model is still able to identify sharp light curves in this  data situation. This is also a consequence of the short duration of KN-like events combined with the ZTF cadence, which generates a sample of KN light curves with a fairly low number of informative light-curve points for training. Although there are still differences between the catalogue level simulations that compose \texttt{ztf\_sims} and the alert photometry experience by \fink, this result indicates that our parameter space and associated classifier are identifying the correct region of the parameter space populated by short-lived, sharp light-curve transients.

\section{Deployment in Fink}
\label{sec:fink}

Based on the method described above, we deployed a science module in \fink broker\footnote{This module is available at \url{https://github.com/astrolabsoftware/fink-science}.}. \fink\ has been processing the ZTF public alert stream since 2019, and it already had a set of science modules focussing on Solar System objects, SNe, or microlensing events, for example. This new module fills the gap for fast transients, opening a new window on the stream, especially for multi-messenger analyses.

We used the medium light-curve baseline described in Section~\ref{subsec:partial_lc}, with the additional step where ZTF magnitudes were converted into \texttt{FLUXCAL} units\footnote{\url{https://github.com/astrolabsoftware/fink-utils/blob/main/fink_utils/photometry/conversion.py}}  before feature extraction. This module was deployed in March 2021, but for completeness, we report here the results obtained from processing data streamed\footnote{For a complete list of the information contained in each alert, see \url{https://zwickytransientfacility.github.io/ztf-avro-alert/schema.html}.} between November 2019 and December 2021 (524 observing nights).   Similarly to other science modules, we set a number of conditions that must be satisfied before alerts are submitted to classification. For this particular module, it is required that: at least two epochs be observed per filter; at most 20 days be between the first alert emitted and the last; at most 20 measurements be between the first alert emitted and the last;
and no galactic counterpart be from the SIMBAD database \citep{SIMBAD:2000}.

Moreover, in order to tag an alert as a KN candidate, our additional set of three criteria is that: the KN score be larger than 0.5 from the module described in this work; the score be higher than 0.5 from the RealBogus algorithm \citep{2019PASP..131c8002M, 2019MNRAS.489.3582D}; and the star-galaxy classification score from SExtractor be above 0.4 \citep{sextractor}.

\noindent Over the 524 ZTF observing nights considered in this work, 1,996 alerts were classified as KNe (1,251 unique objects), which corresponds to about four alerts per night.

\begin{figure}[!t]
  \includegraphics[width=\linewidth]{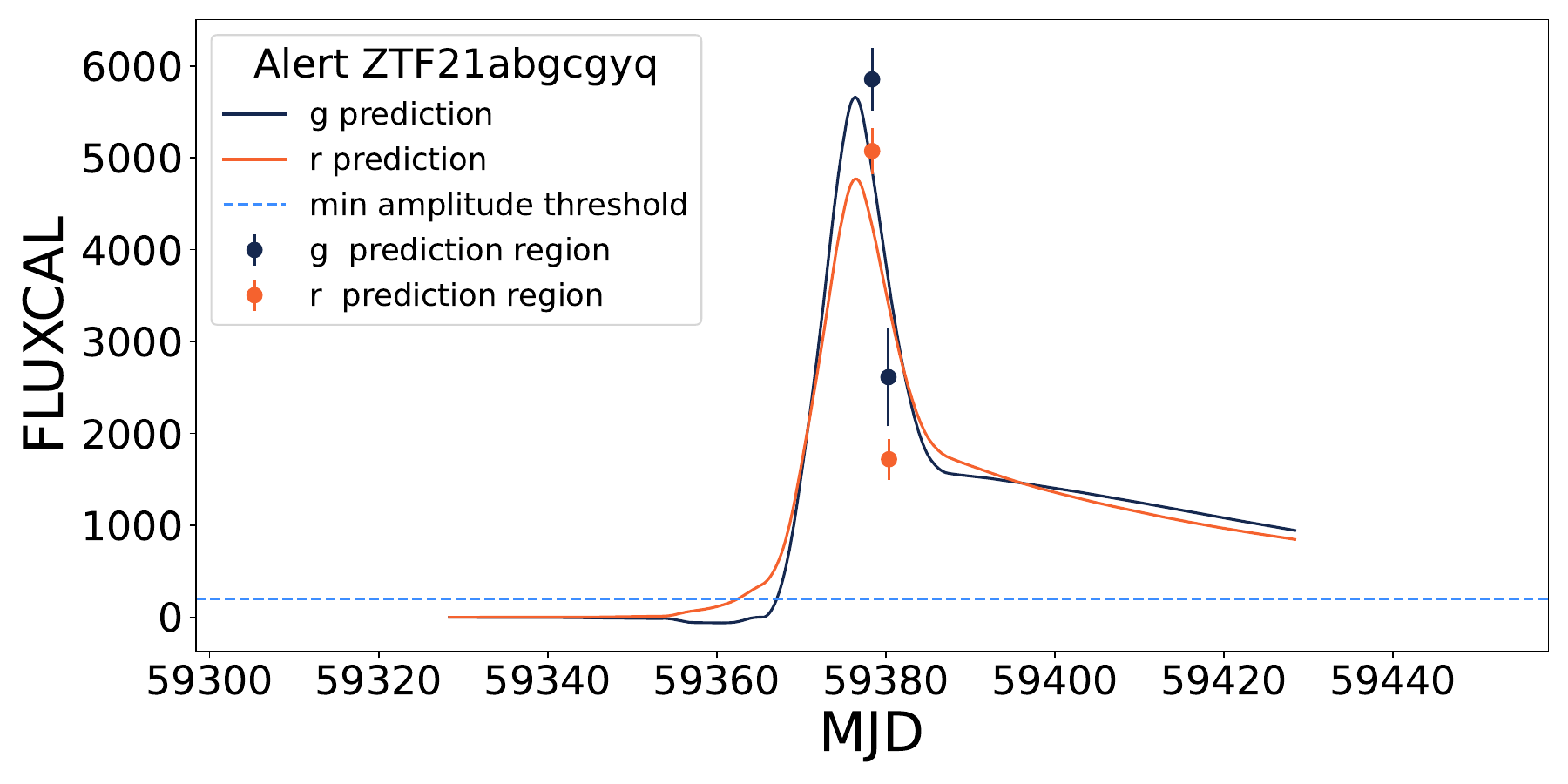}
  \caption{Plot for object ZTF21abgcgyq that was classified as a KN candidate by the module (\protect\url{https://fink-portal.org/ZTF21abgcgyq}, alert ID \texttt{1626368700115015001}).}
  \label{fig: real alert plot}
\end{figure}

Immediately after the alert classification, the KN candidates are publicly available to the community via the \fink\ Livestream service\footnote{Users can easily subscribe to \fink\ output streams via a dedicated client \url{https://github.com/astrolabsoftware/fink-client}. The stream reporting KN candidates is named \texttt{fink\_kn\_candidates\_ztf}. Candidates are also available from the Science Portal at \url{https://fink-portal.org}.}. Figure \ref{fig: real alert plot} shows an example candidate identified within the ZTF alert stream. The candidate rate of a few alerts per night is low enough to enable human inspection before any follow-up decision is made. This was particularly important for the 2021 observing campaign coordinated by the GRANDMA telescope network \citep{10.1093/mnras/stac2054}, as the users of the generated sub-stream can veto events, or on the contrary increase the priority of a certain event prior to any follow-up decisions being made.

During the observing campaign, six KN candidates reported by \fink\ were followed up on by GRANDMA, including instruments from both professional and amateur astronomers, but no confirmed KNe were discovered. At the time of emission, there was little external information about a new KN candidate, but some of them later benefited from spectroscopic follow-up that helped disentangle the nature of the event. One of the KN candidates\footnote{\url{https://fink-portal.org/ZTF21ablssud}} (ZTF21ablssud) was identified as being a cataclysmic variable \citep[see Figure 9 of ][]{aivazyan2022}.  
We gathered available spectroscopic information from the Transient Name Server\footnote{\url{https://www.wis-tns.org/}} (TNS) database, and crossmatched it with the KN candidates. Among the 1,251 objects classified as KNe, 956 (76\%) had no counterparts in TNS (that is no follow-up was reported), 205 (16\%) were classified as SN Ia, 39 (3\%) as SN II, 17 (1\%) as cataclysmic variables, one as a tidal disruptive event, and the rest as other various SN types. We note that some of the objects first classified  as KNe by \fink\ were classified later as SNe (Ia and other types) as more alerts were emitted. Readers can refer to \cite{10.1093/mnras/stac2054} for a discussion about the classification evolution of KN candidates.

\section{Conclusions}
\label{sec:conclusions}

The potential of multi-messenger astronomy will require coordination among many different elements to ensure we will optimally exploit the scientific potential of the incoming dataset. In this context, it is paramount to develop machine-learning methods with physically informed features in order to decrease the volume of data that requires human screening. In this stage, simulations can play an important role in guiding initial searches for rare events, whose amount of observed data is still very limited. 

In this work, we present one possible combination of these two important factors by developing a feature extraction method based on perfect simulated light curves. It can be used to select fast transients, including KNe, among others. This is the core of the KN module currently running within the \fink\ broker and it has already produced interesting candidates, which have been scrutinised by the astronomical community of both professional and amateur astronomers. 
Additionally, the way the model has been trained provides the possibility to incorporate feedback from the community to retrain the model. For example, we could re-run the active learning loop to train the model with correct labels from real spectroscopic observations.

Despite the promising results presented in this work, we emphasise that this light-curve-based classifier is just one stage of a complex system that will certainly lead us to more fast transient detections. In the future, we should also add the crossmatch  to this system with alert streams from other wavelengths, which will highly increase our chances of detecting rare fast transients, such as KNe. 
Given its expected increase in sensitivity and data volume, it is reasonable to expect that the likelihood of finding such transients within LSST will be much higher than the current odds with ZTF.
Moreover, LSST will have six broad-band filters which will cover a larger fraction of the wavelength spectrum than the current two filters available in ZTF, and thus we will have to adapt the model to this new situation by generating templates for each band separately, for example.
Efforts such as the one described in this paper will be paramount to guide our future classifiers and enable discoveries.

\begin{acknowledgements}
We deeply thank the GRANDMA team for their support and follow-up, and especially Cosmin Stachie for providing a set of simulated KN data used to train the model, as well as Sarah Antier and Michael Coughlin for early feedback and discussions that led to the improvement of the science module. We thank David O. Jones for making the \texttt{perfect\_sims} dataset available for this project. This work was developed within the \fink\ community and made use of the \fink\ community broker resources. \fink\ is supported by LSST-France and CNRS/IN2P3. This project has received funding from the European Union’s Horizon 2020 research and innovation program under the Marie Skłodowska-Curie grant agreement No 945304 – Cofund AI4theSciences hosted by PSL University. E.E.O.I.  received financial support from CNRS International Emerging Actions under the project \textit{Real-time analysis of astronomical data for the Legacy Survey of Space and Time} during 2021-2022. M. V. P. acknowledges the funding  by RFBR and CNRS according to the research project No 21-52-15024.

\end{acknowledgements}

%
%

\bibliographystyle{aa}
\bibliography{ref}

\end{document}